\documentclass[a4paper,11pt]{article}
\pdfoutput=1
\usepackage{jcappub}
\usepackage{graphicx}
\graphicspath{{figures/}}
 \usepackage{caption}
\usepackage{mathrsfs}
\usepackage{amssymb}
\usepackage{textcomp}
\usepackage{amsmath, float}
 \usepackage{times}
\usepackage{color}
\usepackage{enumitem}

\title{\boldmath Revisiting dark energy models using differential ages of galaxies}

\author[a,1]{Nisha Rani,\note{Corresponding author.}}
\author[b]{Deepak Jain,}
\author[a]{Shobhit Mahajan,}
\author[a]{Amitabha Mukherjee}
\author[c,d]{and Marek Biesiada}

\affiliation[a]{Department of Physics and Astrophysics, University of Delhi, Delhi 110007, India}
\affiliation[b]{Deen Dayal Upadhyaya College, University of Delhi,
Sector-3, Dwarka, Delhi 110078, India}
\affiliation[c]{University of Silesia, Institute of Physics, Uniwersytecka 4 PL-40-007 Katowice, Poland}
\affiliation[d]{Department of Astronomy, Beijing Normal University, Beijing 100875, China}

\emailAdd{nisharani3105@gmail.com}
\emailAdd{djain@ddu.du.ac.in}
\emailAdd{shobhit.mahajan@gmail.com}
\emailAdd{amimukh@gmail.com}
\emailAdd{marek.biesiada@us.edu.pl}

\abstract{In this work, we use a test based on the differential ages of galaxies for distinguishing the dark energy models. As proposed by Jimenez and Loeb in \cite{jimenez}, relative ages of galaxies can be used to put constraints on various  cosmological parameters. In the same vein, we reconstruct $H_0dt/dz$ and its derivative ($H_0d^2t/dz^2$) using a model independent technique called {\it{non-parametric smoothing}}. Basically, $dt/dz$ is the change in the age of the object as a function of redshift which is directly linked with the Hubble parameter. Hence for reconstruction of this quantity, we use the most recent $H(z)$ data. Further, we calculate $H_0dt/dz$ and its derivative for several models like  Phantom, Einstein de Sitter (EdS), $\Lambda$CDM,  Chevallier-Polarski-Linder (CPL) parametrization, Jassal-Bagla-Padmanabhan (JBP) parametrization and Feng-Shen-Li-Li (FSLL) parametrization. We check the consistency of these models with the results of reconstruction obtained in a model independent way from the data. It is observed that $H_0dt/dz$ as a tool is not able to distinguish between the $\Lambda$CDM, CPL, JBP and FSLL parametrizations but, as expected, EdS and Phantom models show noticeable deviation from the reconstructed results. Further, the derivative of $H_0dt/dz$ for various dark energy models is more sensitive at low redshift. It is found that the FSLL model is not consistent with the reconstructed results, however, the $\Lambda$CDM model is in concordance with the 3$\sigma$ region of the reconstruction at redshift $z\ge 0.3$.}

\begin{document}
\maketitle
\flushbottom
\section{Introduction}
\label{sec:intro}
There is convincing evidence coming from various observations that the Universe is now undergoing an accelerated expansion \cite{riess, perlmutter}. Despite having very precise data, we are still struggling to understand the nature of  the mechanism responsible for this late-time acceleration. It is  believed that some kind of cosmic fluid with negative pressure, called dark energy, is responsible for this. The equation of state for a barotropic fluid,  dark energy in particular, can be written as $p = \omega \rho$ where $p$ is the pressure and $\rho$ the energy density. Cosmological constant, which can be  understood as the energy of the quantum vacuum, corresponds to $\omega=-1$  and is possibly the  simplest explanation for dark energy \cite{sean}. However, there is no compelling theoretical motivation that stops one from proposing other models of dark energy \cite{frieman, caldwell}. These alternative models usually result in a time varying equation of state. A large number of such dark energy models is available in the literature. Some of them are in good agreement with certain observations and some with the other ones. In order to restrict the space of possible consistent dark energy models, it is very important to filter out the models which fit best with most of the observations. Unlike the majority of approaches we will base our inference on a non-parametric reconstruction approach.

In the literature, both model dependent and model independent (parametric and non-parametric) methods have been used to study dark energy models. In the model dependent approach, most of the information is lost due to the presence of the integrand of $H(z)$ in $d_L$ or $d_A$ relations. Model independent approaches include parametric and non-parametric methods. In the  parametric method, one  usually parametrizes $\omega$ (the quantity that characterizes the evolution of dark energy) in different functional forms, then the maximum likelihood method is  used to constrain  $\omega$ \cite{sne, sne1, sne2, sne3, dragan, linder,george}. Recently, Jing-Zhao Qi et al used $Om$ diagnostic technique to distinguish dark energy models \cite{jing}. Celia Escamilla-Rivera also use six different parametrizations of $\omega$ with recent JLA and BAO data to explore which parametrization receives the strongest support from the data \cite{celia}. However, this method may also be biased due to its dependence on the assumed form of parametrization. In order to check the consistency of the parametrization of the  equation of state of dark energy, it might be good to compare the results from the parametric approach with the reconstructed result obtained from a non-parametric method.

The Non-parametric approach has an advantage over the parametric method since it is more robust. Here we use a smoothing technique called Non-Parametric Smoothing (NPS) to reconstruct the quantity $H_0dt/dz$ and its derivative ($H_0d^2t/dz^2$). In the past, NPS method has already been used to understand the expansion history of the Universe, its age and the properties of the dark energy \cite{armangp, arman}. Wu and Yu (2008) applied this method to the Supernova data to study the cosmic acceleration history \cite{puxun08}. Shafieloo and Clarkson (2010) used this technique to test the FLRW models \cite{chris2010}. This smoothing method is further combined with crossing statistics to obtain the constraints on dark energy \cite{arman2012}. To reconstruct the Hubble expansion, Zhengxiang Li. et al. (2015) used $H(z)$ data with NPS method  \cite{zhengli2015}. Recently, Gonzalez et al (2016) also studied cosmological matter perturbations using this methodology \cite{data}. Further, L'Huillier and Shafieloo (2016) implemented this technique to test the FLRW metric and non-local measurement of $H_0$ \cite{arman16}.

In this paper, our focus is on reconstructing $H_0dt/dz$ and its derivative using the above mentioned NPS technique. The approach of studying $H_0dt/dz$ and its derivative is known as the  {\it{differential age method}} and was first introduced by Jimenez and Loeb (2002) \cite{jimenez}. They proposed it to measure the relative ages of passively evolving galaxies and subsequently use them to constrain cosmological parameters. They also claimed that $H_0dt/dz$ and $H_0d^2t/dz^2$ have a  better sensitivity with respect to the change in $\omega$ with $z$, than luminosity distance ($d_L$). Hence, it can provide better constraints on the behaviour of dark energy models. This method encounters a fundamental difficulty, since it is hard to find a pair of passively evolving galaxies and date them accurately. In particular it is not possible to completely disentangle the $H_0dt/dz$ from the effect of metallicity evolution. However, as pointed out by Jimenez \cite{jimenez2003},  it is free from many cosmological systematics inherent to other methods. In 2015, Melia and McClintock used cosmic chronometers data to compare the coasting and $\Lambda$CDM model \cite{meliacc2015}. Rafael et al. also used data from cosmic chronometers to constrain the cosmological scenario where the dark matter sector and dark energy interact directly \cite{rafael2016}. In two more recent papers \cite{Ding2015} and \cite{Zheng2016}, the authors performed an $Om(z)$ test introduced in \cite{Sahni2008} using a sample of 29 $H(z)$ measurements from BAO and cosmic chronometers with the conclusion that there is a tension between $H(z)$ data and the $\Lambda$CDM model.

The quantities we are interested to reconstruct (i.e. $H_0dt/dz$ and its derivative) are not directly observable, so we use the $H(z)$ data to derive them and the error propagation equation to assess the corresponding uncertainty. Then we apply NPS to reconstruct $H_0dt/dz$ as a function of redshift $z$. In order to calculate its derivative, we fit the smooth data of $H_0dt/dz$ with a polynomial whose derivative is used to assess $H_0d^2t/dz^2$ and its corresponding uncertainty. Next, we calculate $H_0dt/dz$ and its derivative for different dark energy models. Since the smoothing process relies solely on the observed data without prior assumptions concerning the Cosmology, the reconstructed result can be used to differentiate the dark energy models.

The paper is organized as follows. Section 2 includes the details of the data used in this work. The methodology is explained in Section 3. Finally in Section 4, we discuss the results from analysis using different dark energy models.

\section{Dataset}
\label{sec:data}
We use the recent Hubble function ($H(z)$) data consisting of $38$ data points \cite{datameng, datamoresco}. $30$ $H(z)$ measurements comes from cosmic chronometers, i.e. massive, early-type galaxies evolving passively on a timescale longer than their age difference. Certain features of their spectra, such as $D4000$ break at $4000 \;{\AA}$ indicative of the evolution of their stellar populations enable us to measure age difference of such galaxies  \cite{jimenez}. In the BAO approach, $H(z)$ measurement is done by using the peak position of the BAO in the radial direction as the standard ruler. Sometimes this approach is known as clustering. In particular we used $8$ data points measured using the clustering (BAO) approach  \cite{baohz,baohz1,baohz2,baohz3,baohz4}. The data is given in Table 1.

\section{Methodology}
\label{sec:method}

The Hubble parameter, $H(z)$ for a flat universe ($\Omega \equiv \Omega_m + \Omega_{DE} = 1$) can be written as

\begin{equation}
\label{eq:hubble}
H(z)=H_0[\Omega_m(1+z)^3+(1-\Omega_m)(1+z)^{3(1+\omega)}]^{1/2}
\end{equation}
where $\Omega_m$ is the matter density and $\Omega_{DE}$ is the dark energy density. Recall that $\omega = p/\rho$.
In popular approaches like the one using SNIa as standard candles, one has to consider the luminosity distance:
\begin{equation}
\label{eq:eos}
d_L(z)=c(1+z)\int\limits_0^z \frac{dx}{H_0[\Omega_m(1+x)^3+(1-\Omega_m)(1+x)^{3(1+\omega)}]^{1/2}}
\end{equation}
However in such an approach, a lot of the information is lost when constraining $\omega$ because $d_L$ is related to the equation of state parameter through an integrand. As mentioned in the Introduction, Jimenez and Loeb introduced the  differential age method in which change of the age of the galaxies with redshift is related to the  Hubble parameter by the following equation
\begin{equation}
\label{eq:define}
\frac{dt}{dz}=-\frac{1}{H(z)(1+z)}
\end{equation}
It is also convenient to use the non-dimensional expansion rate $E(z)=\frac{H(z)}{H_0}$.
Let us introduce the quantity:
\begin{equation}
\label{eq:rename}
P(z) =-\frac{1}{(1+z)E(z)}=H_0\frac{dt}{dz}
\end{equation}
It is $P(z)$ that we would like to reconstruct using the non-parametric smoothing. For this purpose, we define

\begin{equation}
\label{eqn:smoothing}
 P_{m+1}^s(z_i, \Delta)=P_m^s(z_i)+N(z_i) \sum_{j=1}^n{[P^{obs}(z_j)-P^s_m(z_j)]}\it{K}(z_i,z_j)
\end{equation}
For $m=0$, the $P_m^s(z_i)$ on the right hand side of Eq. \eqref{eqn:smoothing}, represents the guess model values. In order to use the smoothing process we need  a guess model to initialize. In our case, we take the flat $\Lambda$CDM Universe as the guess model.
\begin{equation}
\label{eq:guess}
 P^s_0(z_i)=\frac{-1}{(1+z_i)[\Omega_{m0}(1+z_i)^3+(1-\Omega_{m0})]^{1/2}}
\end{equation}
We use  $\Omega_{m0}=0.308$ from the recent Planck result \cite{planck}. In Eq. \eqref{eqn:smoothing}, $m$ and $n$ are integers which represent the number of iterations and the sample size respectively. Here $P^{obs}(z)$ refers to values of $P$ calculated from the data at $z_i, H^{obs}(z_i)$ and $\sigma_{H}(z_i)$. The data is given in Table 1.

\begin{equation}
\label{eq:pob}
 P^{obs}(z_i)=\frac{-1}{(1+z_i)E^{obs}(z_i)}
\end{equation}
The error associated with $P^{obs}(z)$ i.e. $\sigma_P(z)$ is calculated using error propagation equation
\begin{equation}
\label{eq:errorpob}
\sigma_P^2(z_i)=P^2\left[\frac{\sigma_{E(z_i)}^2}{E(z_i)^2} \right]
\end{equation}

To see the effect of the choice of the guess model on the smoothing process, we repeat this smoothing method with two other guess models: Einstein de Sitter and Coasting. We keep the number of iterations same for all models, i.e. $m=25$ and $\Delta=0.72$. The rationale behind such a choice will be explained in the next paragraph. We find that since this is an iterative process, the reconstructed result is almost independent of the choice of the guess model. The plot below shows the reconstructed $H_0dt/dz$ or $P^s(z_i)$ for these three models. It is clear from the graph that reconstructed result does not depend strongly  on the choice of the guess model.

\begin{figure}[ht]
\centering
\includegraphics[width=12.5cm, height=8.5cm]{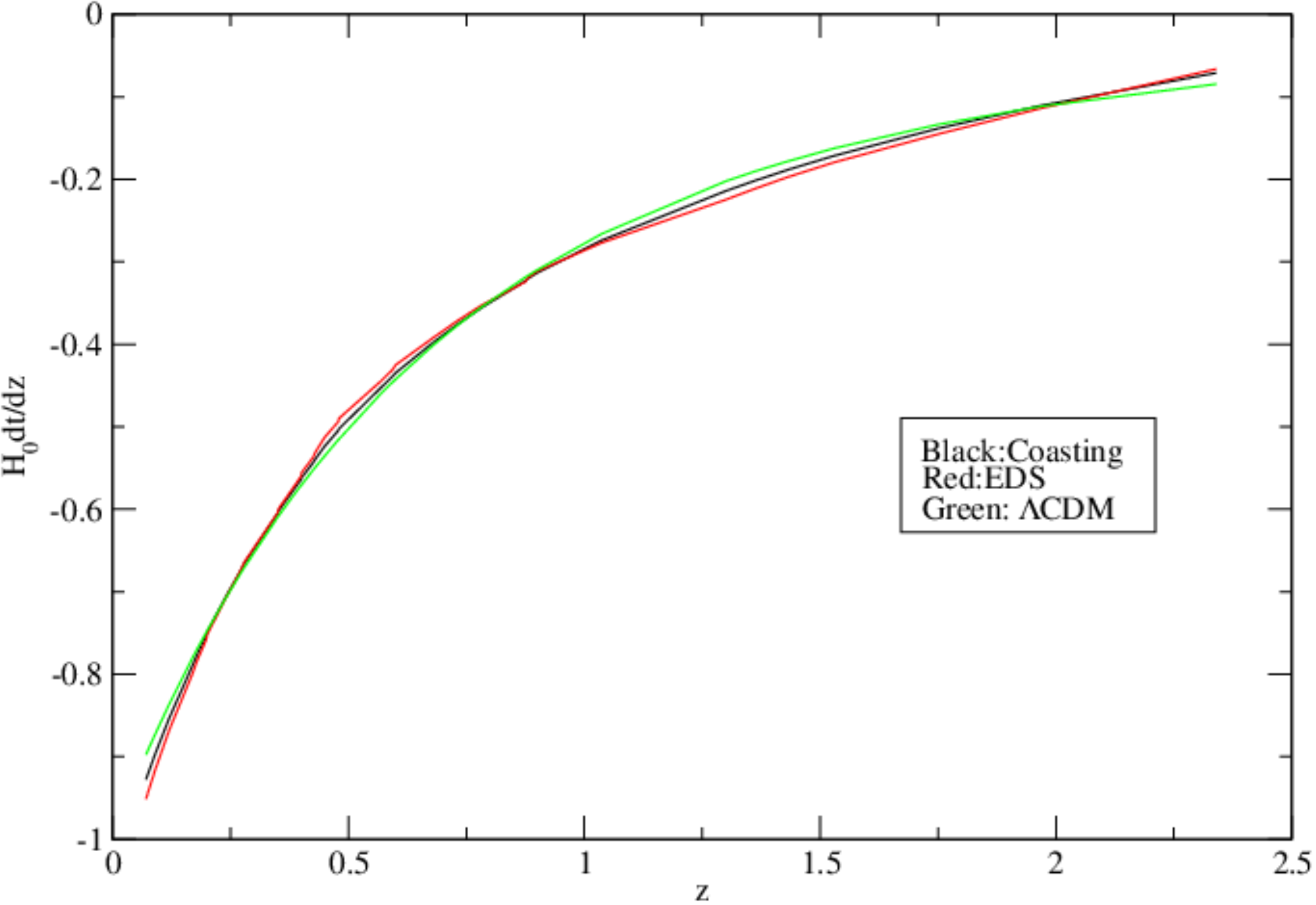}
\caption{\label{fig:diffmodel} Variation of $H_0dt/dz$ (NPS reconstructed) with $z$ for different guess models i.e. $\Lambda$CDM (green), Einstein de Sitter (red) and Coasting model (black).}
\end{figure}
In Eq. \eqref{eqn:smoothing}, $N(z)$ is a normalization factor and can be defined as
\begin{equation}
\label{eq:normalization}
 N(z_i)= \left( \sum_{j=1}^n \it{K}(z_i,z_j) \right)^{-1}
\end{equation}

The kernel $\it{K}(z_i,z_j)$ is defined as
 \begin{equation}
 \label{eq:kernal}
 \it{K}(z_i,z_j)= \exp\left( -\frac{\ln^2(\frac{1+z_j}{1+z_i})}{2 \Delta^2} \right)
 \end{equation}
and $\Delta$ is the smoothing input parameter which is to be fixed. We follow Shafieloo et al (2007) to find the optimal value of $\Delta$ \cite{arman}.
We start by defining $\Delta$ as

\begin{equation}
\label{delta}
\Delta=\sqrt{m} \Delta_0
\end{equation}
where $m$ is the number of iterations. $\Delta_0$ is calculated using the following equation
\begin{equation}
\Delta_0=\left(\frac{1}{3} \right)^{2/3}n^{-1/3}
\end{equation}
where $n$ is number of data points. In the sample we use, $n=38$ which gives  $\Delta_0=0.143$\\
Since $\Delta$ depends on the number of iterations ($m$), for a given $\Delta_0$, the value of $\Delta$ can be different depending upon the choice of $m$. We next turn to a method to determine the appropriate value of $m$.

\begin{figure}[ht]
\centering
\includegraphics[width=12.5cm, height=8.5cm]{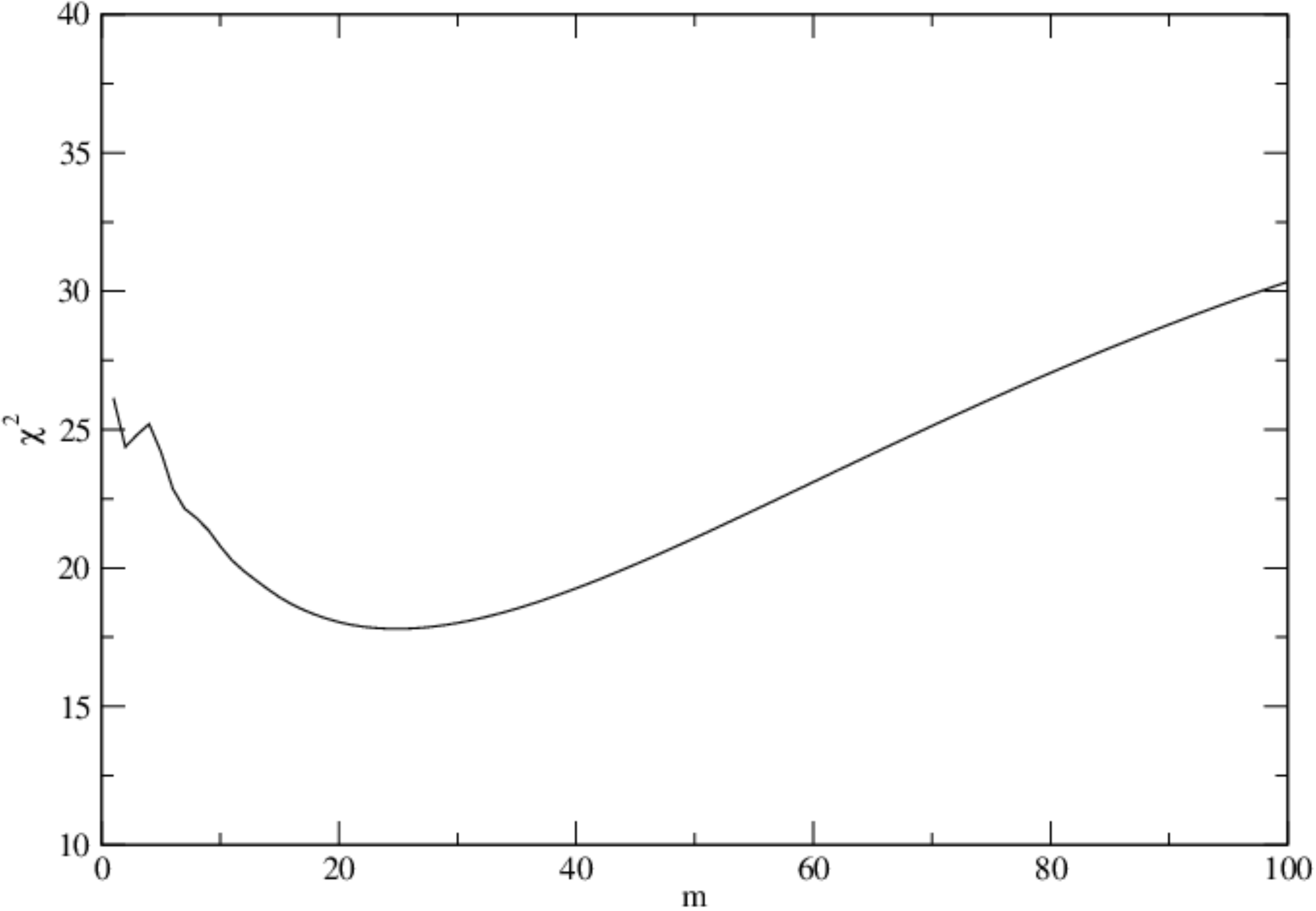}
\caption{\label{fig:chivsm} Variation of $\chi^2$ with iteration (m) in Non-Parametric Smoothing process.}
\end{figure}

To decide the appropriate value of $m$, we start the smoothing process with some random $\Delta$. Here we take $\Delta$ to be $0.32$. With this value of $\Delta$, we calculate the $\chi^2$ after each iteration of smoothing process upto $100$ iterations which is defined as:

 \begin{equation}
 \label{chisq}
 \chi^2_m=\sum_{i=1}^n\frac{[P^s_m(z_i)-P^{obs}(z_i)]^2}{\sigma^2_{{P}^{obs}}(z_i)}
 \end{equation}
To calculate $\chi^2_1$ we use Eq. \eqref{chisq}. $P_1^s(z_i)$ is calculated by using $P_0^s(z_i)$, $N(z_i)$ and $\it{K}(z_i,z_j)$ [see Eq. \eqref{eqn:smoothing}]. In the next iteration, we replace guess model values i.e. $P_0^s(z_i)$ by  $P_1^s(z_i)$ to get $\chi^2_2$ and this process continues till we reach the required number of iterations i.e the iteration at which $\chi^2$ becomes minimum. It is clear from Figure \ref{fig:chivsm} that around $m=25$, $\chi^2$ is minimum so we decide to stop smoothing process at this iteration. For $m=25$, the value of $\Delta$ comes out to be $0.72$ [see Eq. \eqref{delta}] and we consider this value as the optimal value of $\Delta$.

We also checked how $\chi^2$ varies with $m$ for other values of $\Delta$ like 0.64, 0.72, 0.78 and 1.01. But we find that the clear minimum exist only in the case of $\Delta=0.32$. Because we noticed that after $m=25$, the change in $\chi^2$ is very small for other values of $\Delta$ so we consider $m=25$ as the proper iteration to stop the smoothing process.

\begin{figure}[ht]
\centering
\includegraphics[width=12.5cm, height=8.5cm]{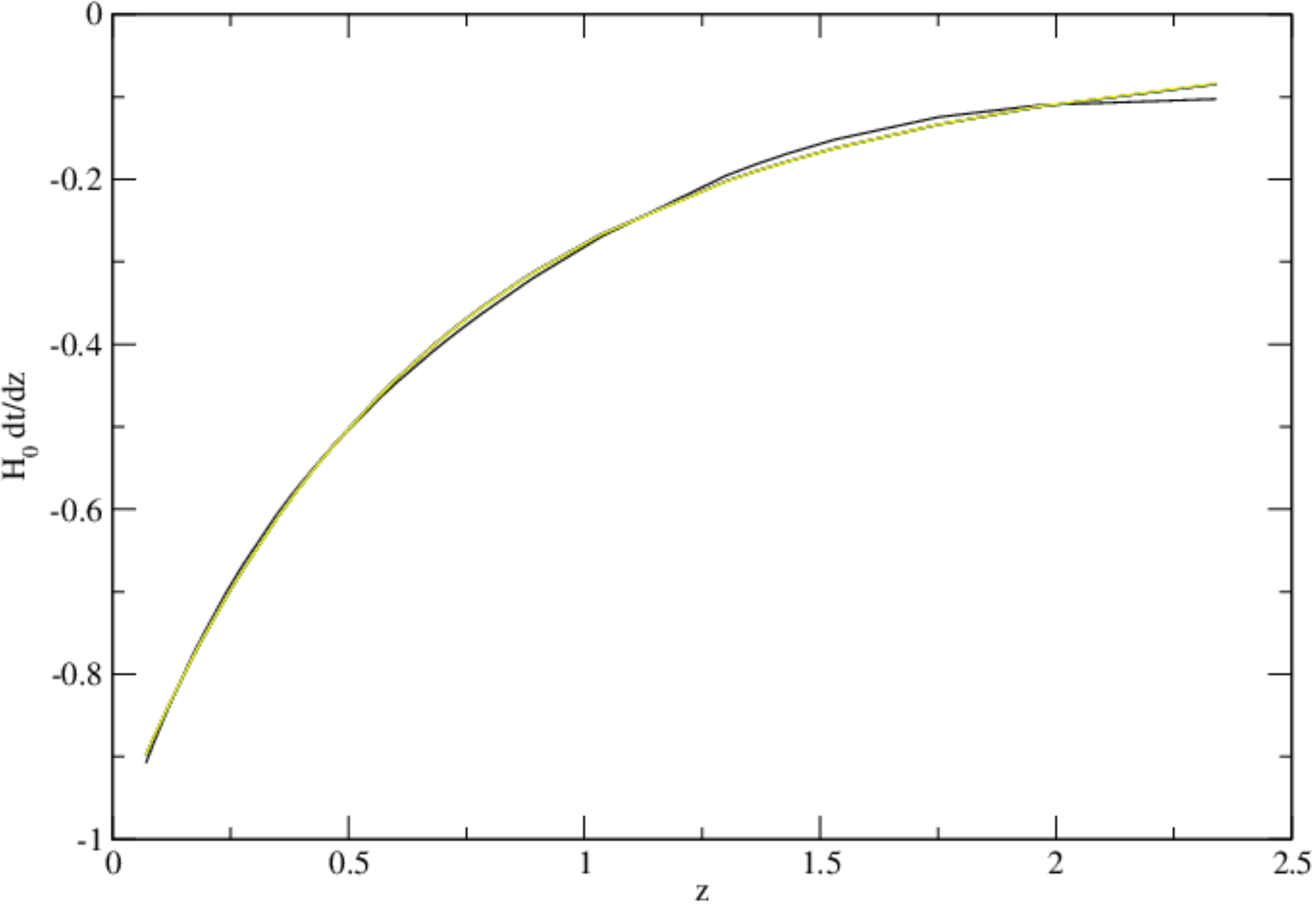}
\caption{\label{fig:diffdelta} Variation of $H_0dt/dz$ (NPS reconstructed) with $z$ for different $\Delta$ values. Black, red, green, blue \& yellow represent $\Delta=0.32, 0.64, 0.72, 0.78, 1.01 $ with corresponding $m=5, 20, 25, 30, 50 $ respectively.}
\end{figure}

For given $m$, a very high value of $\Delta$ can give smooth but less accurate results. On the other hand, small values of $\Delta$ can give accurate but noisy results. So the choice of optimal $\Delta$ and its corresponding $m$ plays a vital role in this process. Figure \ref{fig:diffdelta} shows variation of reconstructed $H_0dt/dz$ with $z$ for some values of $\Delta$ corresponding to different number of iterations ($m$). From this it is clear that if the number of iterations is tuned simultaneously with chosen value of $\Delta$, then the  reconstructed results remain unaffected (see Figure \ref{fig:diffdelta}).

\begin{figure}[ht]
\centering
\includegraphics[width=12.5cm, height=8.5cm]{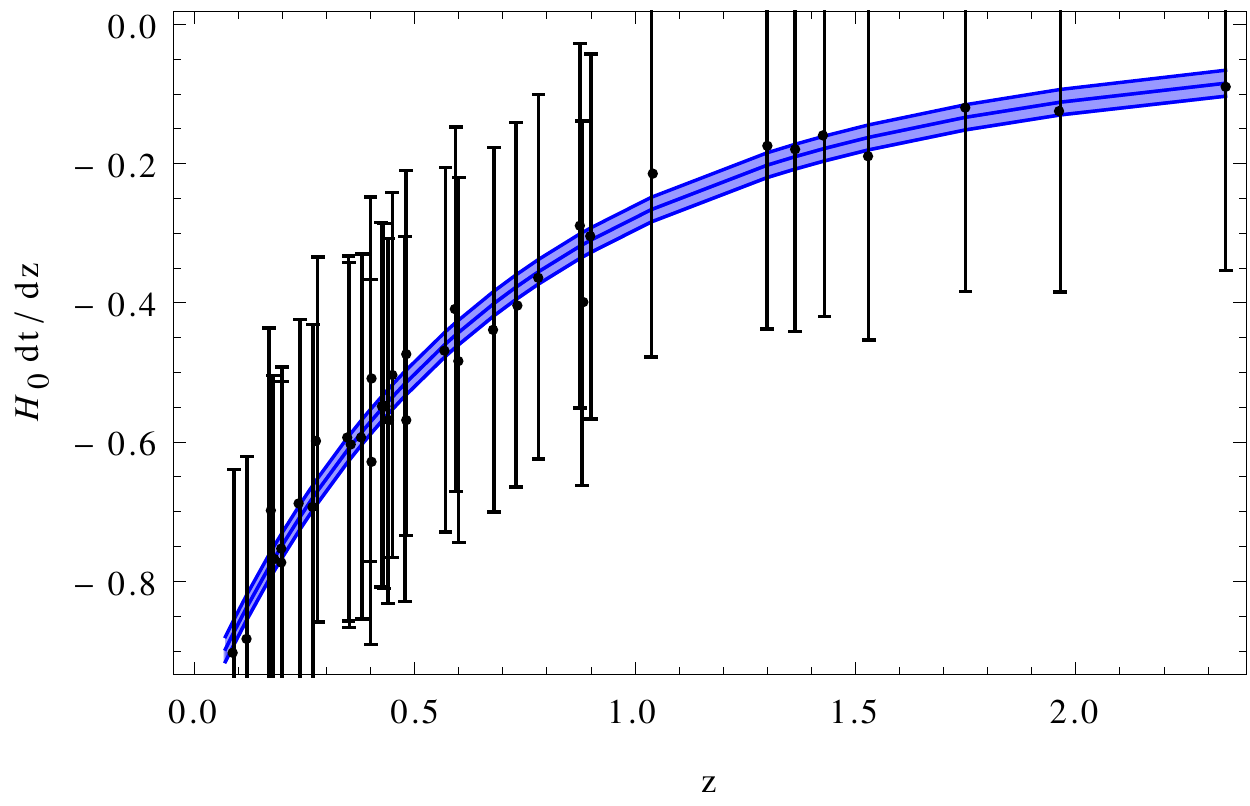}
\caption{\label{fig:dtbydznps}Variation of $H_0dt/dz$ with $z$ and its 3$\sigma$ error bands (blue) reconstructed using non-parametric smoothing. Black points are the observed $H_0dt/dz$ with error bars.}
\end{figure}

In order to calculate the error bands in the reconstructed values, we use the method discussed in Zhengxiang Li et al. \cite{zhengli2015}. We define $\sigma_P^s(z_i)$ as

\begin{equation}
\sigma_P^s(z_i)=\left(\sum_j v_{ij}^2\hat{\sigma_i}^2\right)^{1/2}
\end{equation}
where $\sigma_P^s(z_i)$, $v_{ij}$ and $\hat{\sigma_i}^2$ represent errors in the reconstructed data, the smoothing factor and the estimate of the error variance respectively. Here,

\begin{equation}
v_{ij}=N(z_i)K(z_i,z_j)
\end{equation}
and
 \begin{equation}
 \label{errornps}
 \hat{\sigma_i}^2=\frac{\sum_k [P^{obs}(z_k)-P^s(z_k)]^2}{\sum_k(1-v_{ik})}
 \end{equation}
To calculate the 3$\sigma$ error band, we simply multiply  the 1$\sigma$ error by $3$. In Figure \ref{fig:dtbydznps} we display the complete reconstructed $H_0dt/dz$ vs $z$ plot including the 3$\sigma$ error bands.

\subsection{$H_0\frac{dt}{dz}$ as a tool}
There are many models for dark energy in the literature \cite{demodels,demodels1}.  We focus only on six popular models: Phantom, EdS, $\Lambda$CDM, CPL JBP \& FSLL. Our purpose is to find the models which are in good agreement with the results obtained from the non-parametric smoothing technique.

\subsubsection{Phantom model}
For the Phantom model, the reduced Hubble parameter is given by
\begin{equation}
\label{hzxcdm}
E_{phantom}(z)=\frac{H_{phantom}(z)}{H_0}=\left[\Omega_{m0}(1+z)^3+(1-\Omega_{m0})(1+z)^{3(1+\omega)}\right]^{1/2}
\end{equation}
and
\begin{equation}
\label{pxcdm}
P^{th}_{phantom}(z)=- \frac{1}{(1+z)\left[\Omega_{m0}(1+z)^3+(1-\Omega_{m0})(1+z)^{3(1+\omega)}\right]^{1/2}}
\end{equation}

In this model, $\Omega_{m0}$ is a free parameters. We fix $\omega$ to be -2 and take $\Omega_{m0}=0.308\pm0.012$ from the Planck result \cite{planck}. The $\pm 1\sigma$ errors in $P_{phantom}^{th}$ are calculated using following Eq. \cite{error1, error2}

\begin{equation}
\label{sglcdmp}
\sigma_{P+}=\sqrt{ \sum_i\left(Max \left[\frac{dP}{dx_i}\sigma_{x_{i+}}, -\frac{dP}{dx_i}\sigma_{x_{i-}}\right]\right)^2 }
\end{equation}
\begin{equation}
\label{sglcdmp1}
\sigma_{P+}=\sqrt{ \sum_i\left(Min \left[\frac{dP}{dx_i}\sigma_{x_{i+}}, -\frac{dP}{dx_i}\sigma_{x_{i-}}\right]\right)^2 }
\end{equation}

$\sigma_{x_{i+}}$ and $\sigma_{x_{i-}}$ are the $\pm1\sigma$ errors, where $x_i$ represents free parameters of the model.

\begin{figure}[ht]
\centering
\includegraphics[width=12.5cm, height=8.5cm]{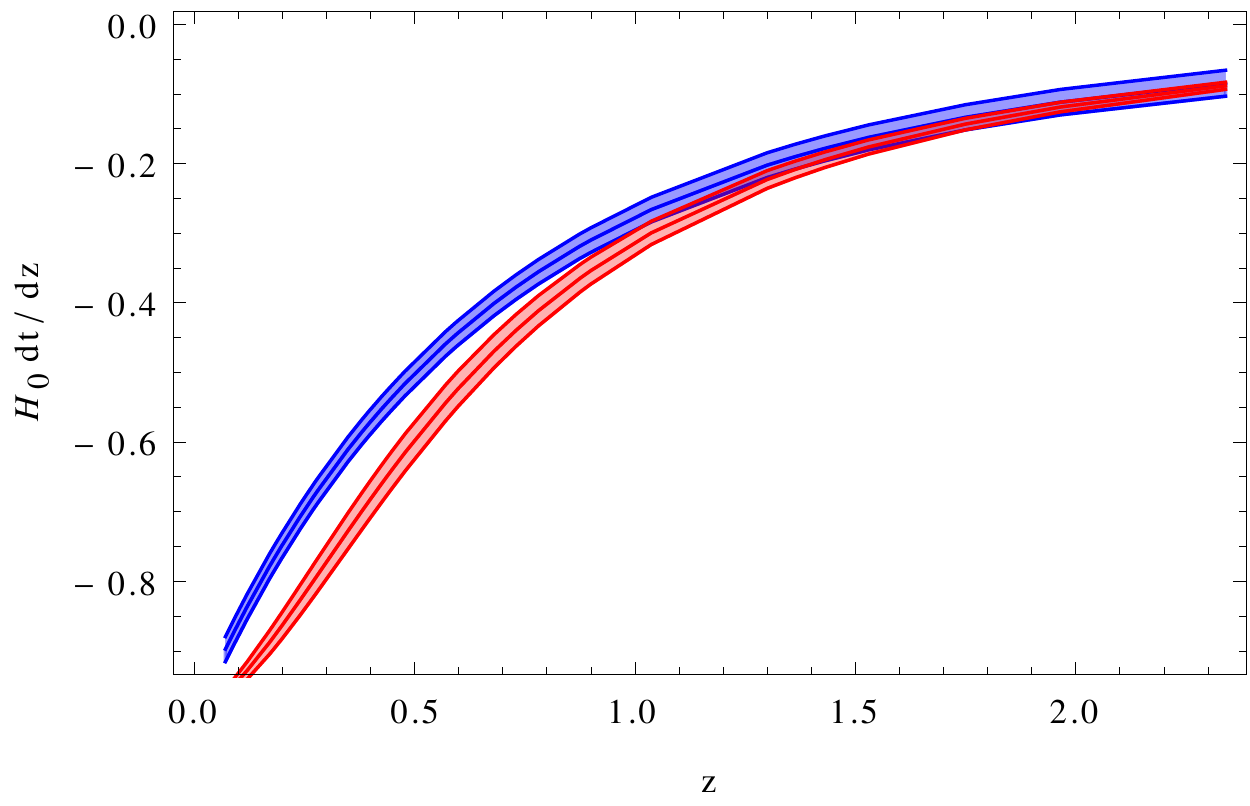}
\caption{\label{dtbydzxcdmwidnps} Variation of $H_0dt/dz$ with 3$\sigma$ error bands for Phantom model (red) and reconstructed result (blue) with $z$.}
\end{figure}

\subsubsection{Einstein de Sitter (EdS) model}

\begin{figure}[ht]
\centering
\includegraphics[width=12.5cm, height=8.5cm]{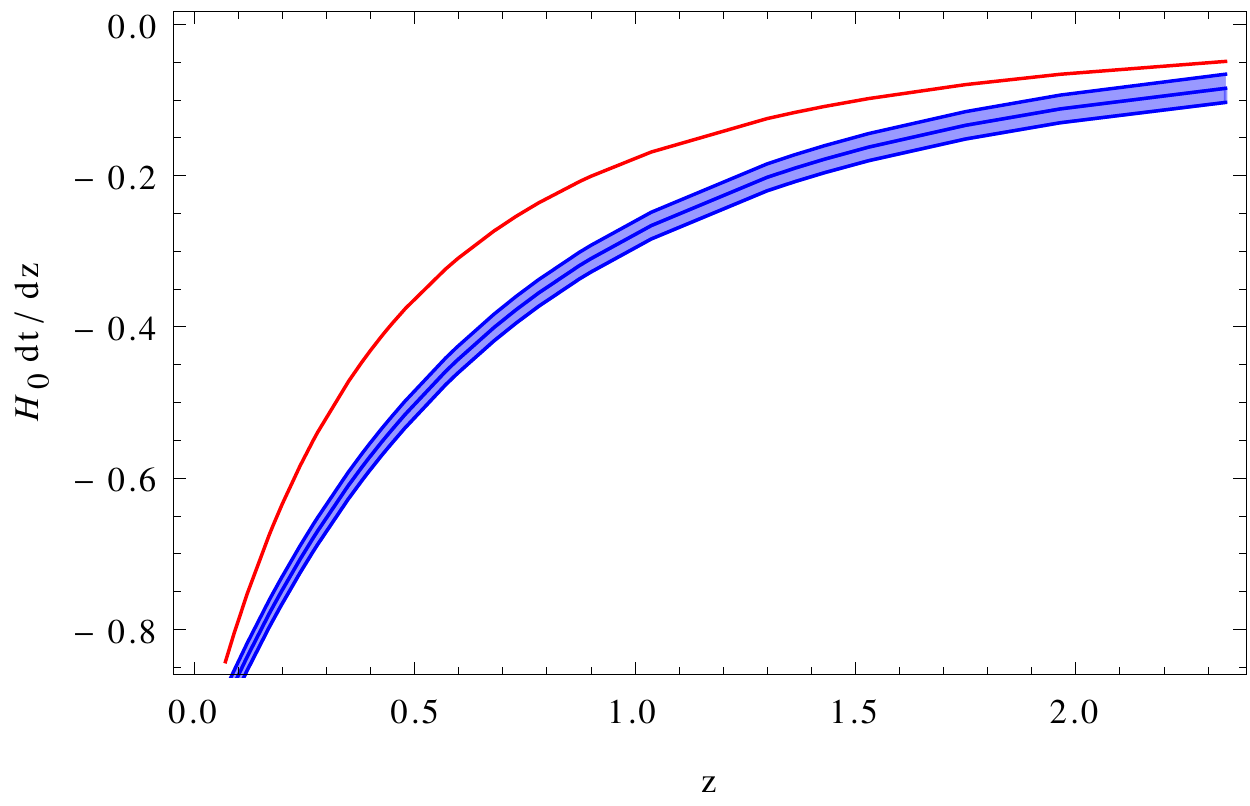}
\caption{\label{dtbydzEdSwidnps}Variation of $H_0dt/dz$ for EdS model (red) and the reconstructed result (blue) with $z$.}
\end{figure}

The dimensionless Hubble parameter for the EdS model is written as:
\begin{equation}
\label{hzEdS}
E_{EdS}(z)=\frac{H_{EdS}(z)}{H_0}=(1+z)^{3/2}
\end{equation}
Correspondingly,
\begin{equation}
\label{dtbydzEdS}
P_{EdS}^{th}(z)=H_0\frac{dt}{dz}=\frac{-1}{(1+z)^{5/2}}
\end{equation}
In this model, there is no free parameter. In a similar way to the Phantom model, we can calculate $P_{EdS}$. The variation of $H_0dt/dz$ with $z$ is shown in Figure \ref{dtbydzEdSwidnps}.

\subsubsection{$\Lambda$CDM model}

\begin{figure}[ht]
\centering
\includegraphics[width=12.5cm, height=8.5cm]{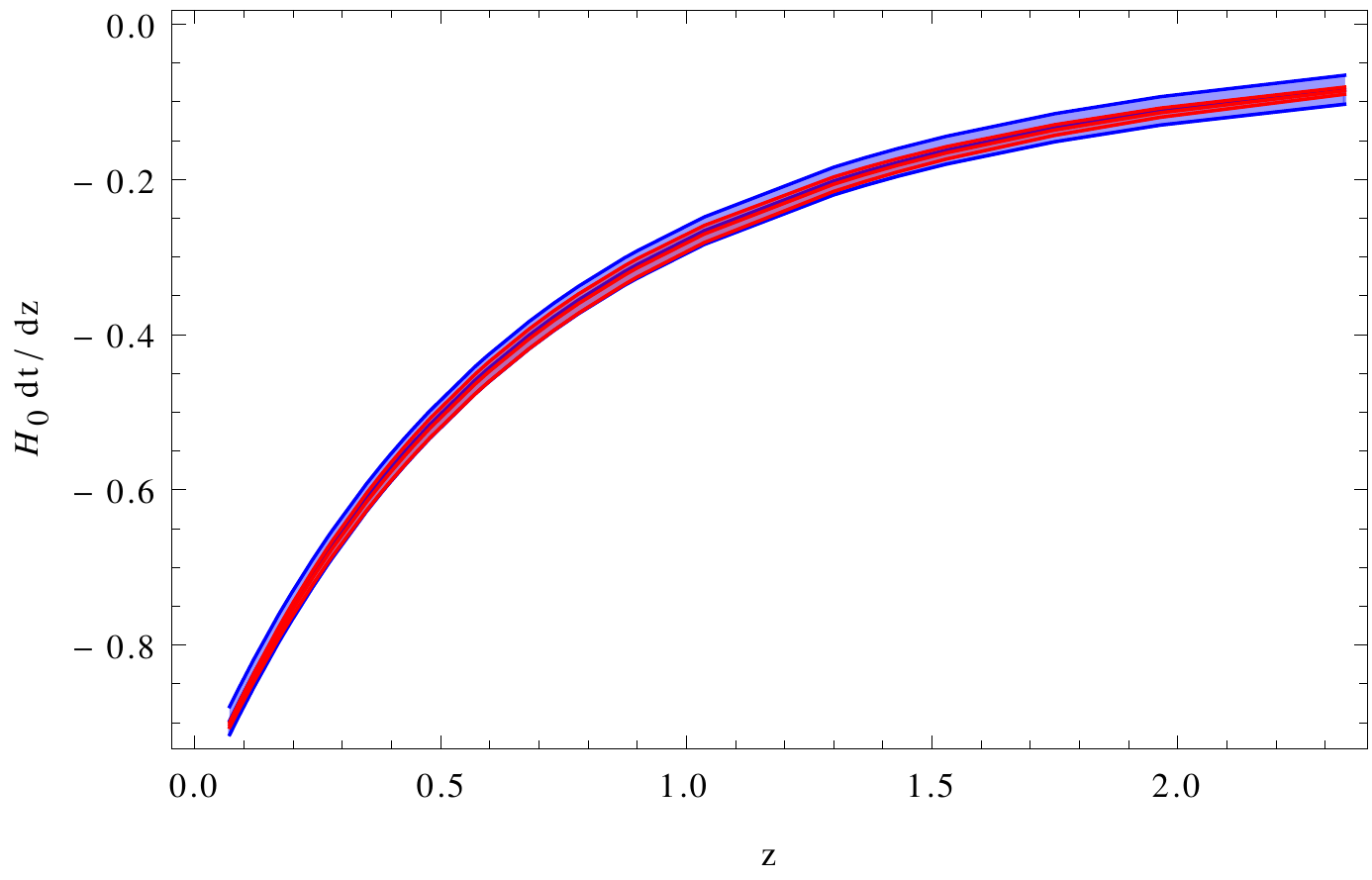}
\caption{\label{dtbydzlcdmwidnps}Variation of $H_0dt/dz$ with $z$ for $\Lambda$CDM model (red) and NPS technique (blue) with corresponding 3$\sigma$ error bands.}
\end{figure}

This model is in very good agreement with most of the observations, hence it is widely accepted. In a flat Universe, the dimensionless Hubble parameter for $\Lambda$CDM model is
\begin{equation}
\label{hubbleflat}
E_{\Lambda CDM}(z)=\frac{H_{\Lambda CDM}(z)}{H_0}=[\Omega_{m0}(1+z)^3+(1-\Omega_{m0})]^{1/2}
\end{equation}
 and therefore the corresponding $P_{\Lambda CDM}$ is given by

\begin{equation}
\label{plcdm}
P_{\Lambda CDM}^{th}(z)=H_0\frac{dt}{dz}=-\frac{1}{(1+z)[\Omega_{m0}(1+z)^3+(1-\Omega_{m0})]^{1/2}}
\end{equation}
Here we have one free parameter, $\Omega_{m0}$.  We take $\Omega_{m0}=0.308\pm0.012$ \cite{planck}. The result obtained is shown in Figure \ref{dtbydzlcdmwidnps}.

\subsubsection{ {\textbf C}hevallier-{\textbf P}olarski-{\textbf L}inder {\textbf (CPL)} Parametrization}
The equation of state for dark energy has been parametrized in many forms. CPL is one of the most popular parametrizations of the dark energy equation of state. It has two free parameters $\omega_0$ and $\omega_1$ and can be  written as \cite{cpl1,cpl2}

\begin{equation}
\label{eoscpl}
\omega_{CPL}=\omega_0+\omega_1(1-a)=\omega_0+\omega_1\frac{z}{1+z}
\end{equation}
and the corresponding Hubble parameter

\begin{equation}
\label{hzcpl}
E_{CPL}(z)=\frac{H_{CPL}(z)}{H_0}=\left[\Omega_{m0}(1+z)^3+(1-\Omega_{m0})(1+z)^{3(1+\omega_0+\omega_1)}\exp\left( - \frac{3\omega_1z}{1+z} \right)\right]^{1/2}
\end{equation}

This gives

\begin{equation}
\label{dtbydzcpl}
P_{CPL}^{th}(z)=\frac{-1}{(1+z)[\Omega_{m0}(1+z)^3+(1-\Omega_{m0})(1+z)^{3(1+\omega_0+\omega_1)}\exp\left( -\frac{3\omega_1z}{1+z} \right)]^{1/2}}
\end{equation}

\begin{figure}[ht]
\centering
\includegraphics[width=12.5cm, height=8.5cm]{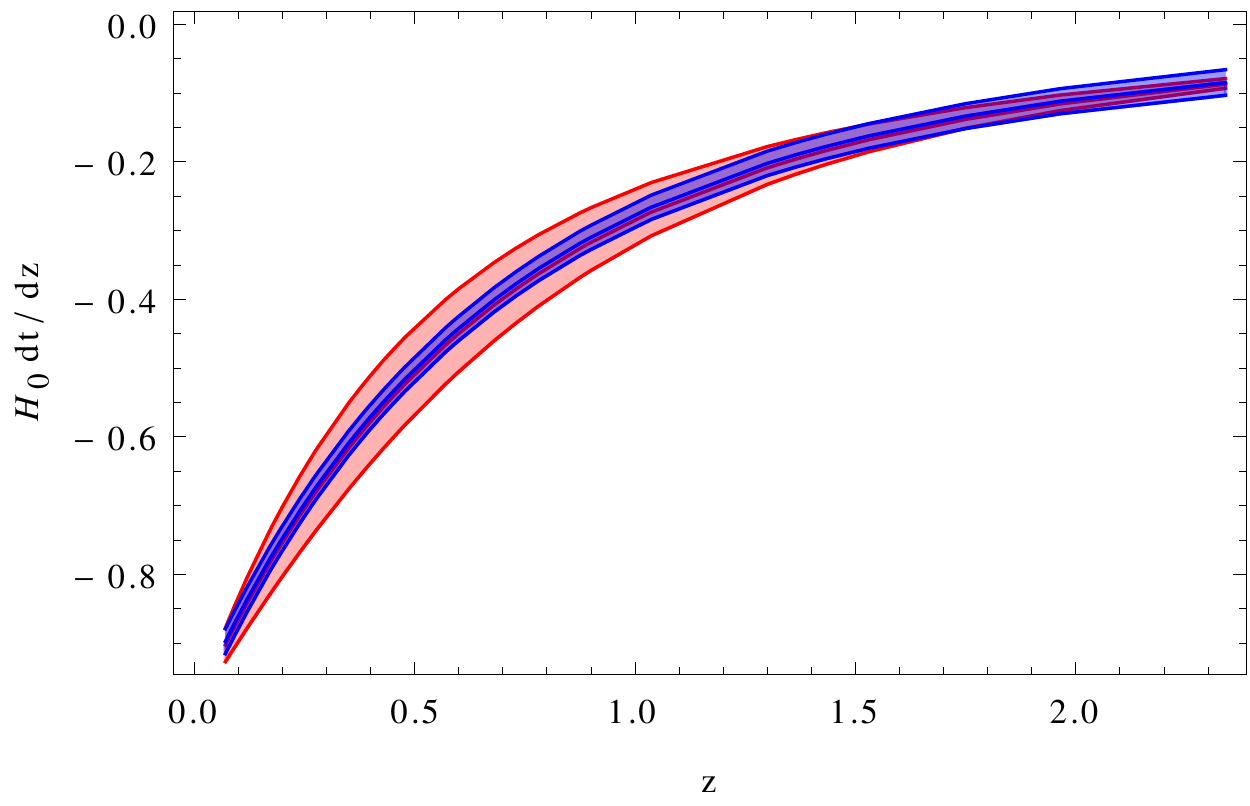}
\caption{\label{dtbydzcplwidnps}Variation of $H_0dt/dz$ and its 3$\sigma$ error bands with $z$ for CPL parametrization of equation of state of dark energy (red) and non-parametric smoothing (blue).}
\end{figure}

We use $\Omega_{m0}=0.300\pm0.0014$, $\omega_0=-0.982\pm0.134$ and $\omega_1=-0.082^{+0.655}_{-0.440}$  obtained  from the joint analysis of Supernovae Ia (SNeIa), Baryonic Acoustic Oscillations (BAO) and Cosmic Microwave Background (CMB) data \cite{wang}. $H_0dt/dz$ variation with $z$ for CPL parametrization is shown in Figure \ref{dtbydzcplwidnps}.

\subsubsection{ {\textbf J}assal-{\textbf B}agla-{\textbf P}admanabhan {\textbf(JBP)} Parametrization}
Equation of state for this parametrization is \cite{jbp}
\begin{equation}
\label{eosjbp}
\omega_{JBP}=\omega_2+\omega_3a(1-a)=\omega_2+\omega_3\frac{z}{(1+z)^2}
\end{equation}
and corresponding reduced Hubble parameter is
\begin{equation}
\label{hzjbp}
E_{JBP}(z)=\frac{H_{JBP}(z)}{H_0}=\left[\Omega_{m0}(1+z)^3+(1-\Omega_{m0})(1+z)^{3(1+\omega_2)}\exp\left(\frac{3\omega_3z^2}{2(1+z)^2} \right) \right]^{1/2}
\end{equation}
and the $P^{th}_{JBP}$ is given by
\begin{equation}
\label{dtbydzjbp}
P^{th}_{JBP}(z)=\frac{-1}{(1+z)\left[\Omega_{m0}(1+z)^3+(1-\Omega_{m0})(1+z)^{3(1+\omega_2)}\exp\left(\frac{3\omega_3z^2}{2(1+z)^2} \right) \right]^{1/2}}
\end{equation}

\begin{figure}[ht]
\centering
\includegraphics[width=12.5cm, height=8.5cm]{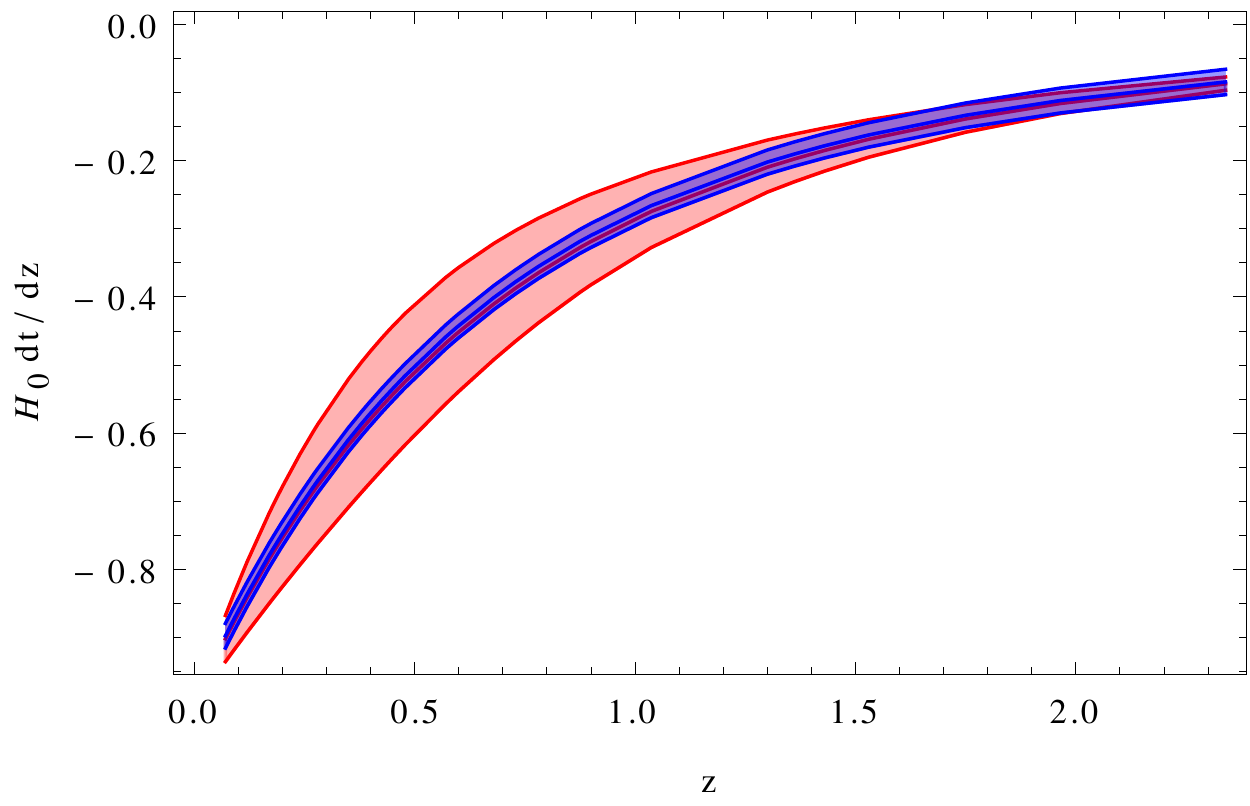}
\caption{\label{dtbydzjbpwidnps}Variation of $H_0dt/dz$ and 3$\sigma$ error bands with $z$ for JBP parametrization (red) and NPS method (blue). }
\end{figure}

We take $\Omega_{m0}=0.298^{+0.013}_{-0.014}$, $\omega_2=-0.960^{+0.181}_{-0.179}$ and $\omega_3=-0.317^{+1.341}_{-1.149}$  from \cite{wang}. Result obtained for this parametrization is shown in Figure \ref{dtbydzjbpwidnps}.

\subsubsection{ {\textbf F}eng-{\textbf S}hen-{\textbf L}i-{\textbf{L}i} {\textbf(FSLL)} Parametrization}

For FSLL parametrization, the equation of state is defined as \cite{fssl}
\begin{equation}
\label{eosfssl}
\omega_{FSLL}=\omega_4+\omega_5\frac{z}{1+z^2}
\end{equation}
and corresponding reduced Hubble parameter is
\begin{equation}
\label{hzfssl}
E_{FSLL}(z)=[\Omega_{m0}(1+z)^3+(1-\Omega_{m0})(1+z)^{3(1+\omega_4-0.5\omega_5)}(1+z^2)^{0.75 \omega_5}\exp(1.5 \omega_5 \tan^{-1}(z)]^{1/2}
\end{equation}
and the $P^{th}_{FSLL}$ is given by
\begin{equation}
\label{dtbydzfssl}
P^{th}_{FSLL}(z)=\frac{-1}{(1+z)E_{FSLL}(z)}
\end{equation}

\begin{figure}[ht]
\centering
\includegraphics[width=12.5cm, height=8.5cm]{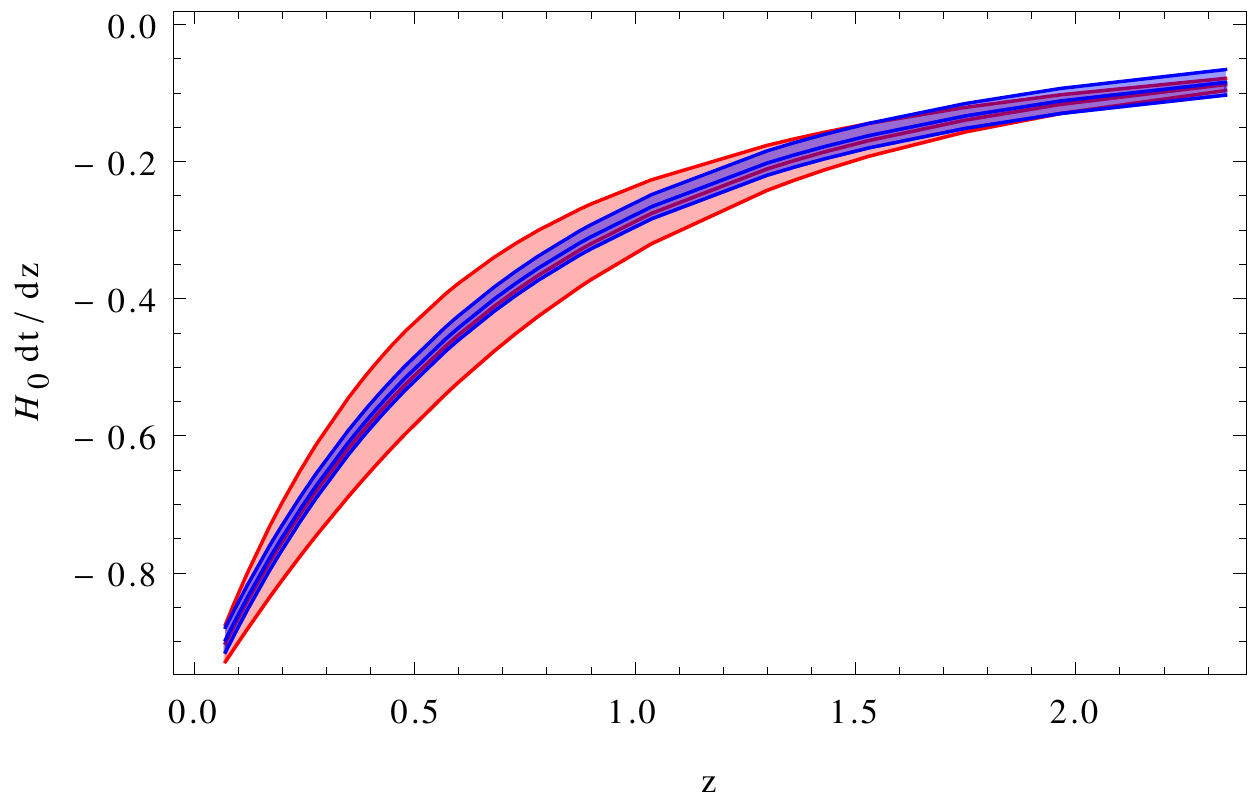}
\caption{\label{dtbydzfsslwidnps}Variation of $H_0dt/dz$ and 3$\sigma$ error bands with $z$ for FSLL parametrization (red) and NPS method (blue).}
\end{figure}

We take $\Omega_{m0}=0.295^{+0.013}_{-0.015}$, $\omega_4=-0.968^{+0.145}_{-0.144}$ and $\omega_5=-0.165^{+0.641}_{-0.527}$  from \cite{wang}. For FSLL parametrization, result is shown in Figure \ref{dtbydzfsslwidnps}.

\begin{figure}[ht]
\centering
\includegraphics[width=12.5cm, height=8.5cm]{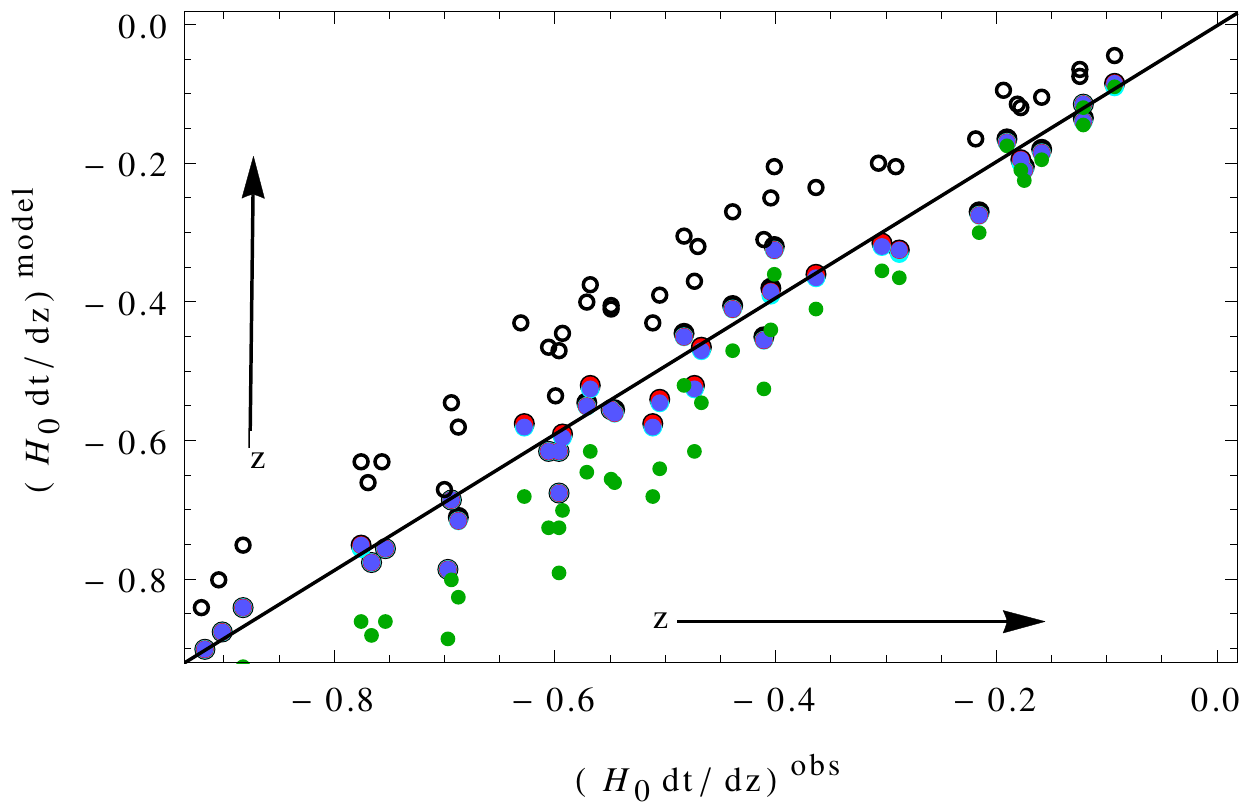}
\caption{\label{dtbydzobsnmodel}Variation of $H_0dt/dz$ for all models with the corresponding observed $H_0dt/dz$. Empty black and filled green circles represents the EDS and phantom model respectively. Filled Black, red, blue and cyan dots represents $\Lambda$CDM, CPL, JBP and FSLL respectively. The arrows indicate the direction of increasing redshift.}
\end{figure}

 The uncertainties of $P^{th}$ for all the models, were calculated using the Eq. \eqref{sglcdmp} and \eqref{sglcdmp1} as explained in section 3.1.1. From Figures \ref{dtbydzlcdmwidnps}, \ref{dtbydzcplwidnps}, \ref{dtbydzjbpwidnps} and \ref{dtbydzfsslwidnps}, it seems that all four dark energy models i.e. $\Lambda$CDM, CPL, JBP and FSLL are consistent with the result obtained from smoothing method. The predictions of  Phantom model is only accommodated within $3\sigma$ region of the reconstructed results at high redshift ($z>1.6$). However, Einstein de Sitter model is inconsistent with the NPS results at all redshifts.
Figure \ref{dtbydzobsnmodel} compares theoretical predictions of the models considered with corresponding observed values of $H_0dt/dz$.
Consistency would be marked by dots following the diagonal black line in the figure. One can see that for the EdS model, theoretical value of $H_0dt/dz$ is always higher than the observed value.
For the phantom model, on the other hand, theoretical predictions are lower than the observed ones. Other four dark energy models ($\Lambda$CDM, CPL, JBP and FSLL), give predictions which are scattered around the diagonal line, which means that they are consistent with the data.

Here we take the NPS reconstructed quantities as representing the true cosmology, in the sense that we did not make any assumptions besides homogeneity, isotropy and spatial flatness of the Universe. We then need to compare it with the predictions of the  six
parametrized models (i.e. depending on parameters like $\Omega_{m0}$, $\omega_i$). Of course there could be some ranges of such parameters for which  the models are compatible with the  NPS reconstruction. The identification of such values is the problem of model fitting, and even when found they would trigger the question whether they are compatible with other probes. Therefore our strategy is to take for comparison the values of cosmological parameters (with their uncertainties) already best fitted to precise measurements from the SN Ia, CMB or BAO. One might be worried theat such an approach will introduce a bias. Indeed, precisely because it is intended to guarantee that the models considered are in agreement with the cosmological probes, the results are biased as long as the data from CMB (Planck data) or SN Ia or BAO are biased.

It is observed that $H_0dt/dz$ is unable to distinguish the four models viz. $\Lambda$CDM, CPL, JBP and FSLL, we follow an idea proposed by  Jimenez and Loeb (2002)  to use the derivative of $H_0dt/dz$ to compare the predictions of these degenerate models with the reconstructed one.

\subsection{Derivative of $H_0dt/dz$}

\begin{figure}[ht]
\centering
\includegraphics[width=12.5cm, height=8.5cm]{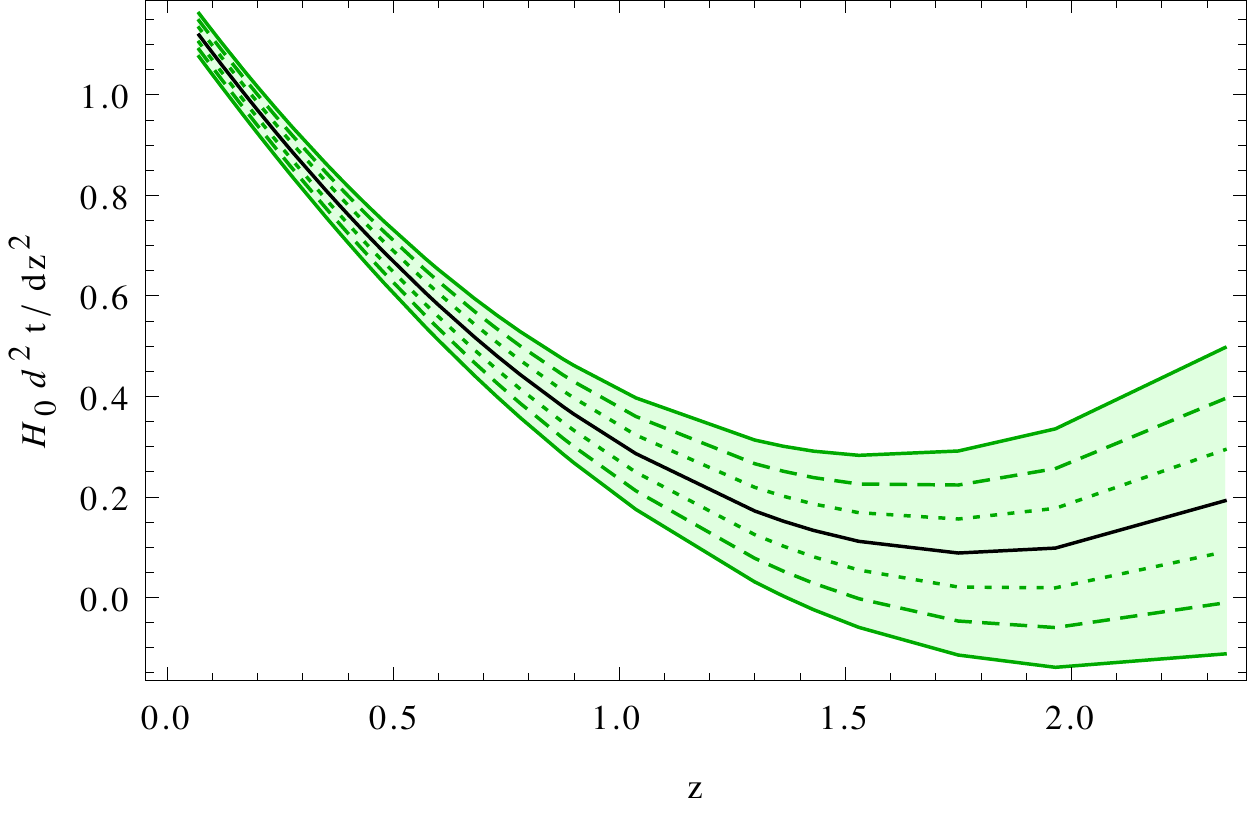}
\caption{\label{d2tbydz2nps}Variation of reconstructed $H_0d^2t/dz^2$ with $z$ where dotted, dashed and continuous lines shows 1$\sigma$, 2$\sigma$, 3$\sigma$ confidence region respectively. The solid black line is the best fit line.}
\end{figure}

From Eq. \eqref{eq:define}
$$P(z)=H_0\frac{dt}{dz}=\frac{-1}{E(z)(1+z)}$$
Differentiating this, we get $H_0d^2t/dz^2$.
\begin{equation}
\label{dtbydzderi}
P'(z)=H_0\frac{d^2t}{dz^2}=\frac{1}{E(z)(1+z)}\left[ \frac{1}{(1+z)}+\frac{E'(z)}{E(z)}\right]
\end{equation}
In order to compare $H_0d^2t/dz^2$ obtained from models with smooth $H_0d^2t/dz^2$, we need to first calculate $(H_0d^2t/dz^2)^s$ [hereafter $P'^s_{fit}$]. To do this we proceed as follows: From the observation data, we have $H_0dt/dz$ i.e. $P^{obs}(z)$ and the corresponding error i.e. $\sigma_{P^{obs}}(z)$, to which we apply the smoothing process to get $P^s(z)$. Hence we obtain $P^s(z)$ corresponding to each redshift where $P^{obs}(z)$ is known.  After that we find a polynomial, say $P^s_{fit}(z)$ [Eq. \eqref{pfit}] which fits best with the smooth $P^s(z)$ values.

\begin{equation}
\label{pfit}
P^s_{fit}(z)=A+Bz+Cz^2+Dz^3
\end{equation}
Here we use the chi-square method to find the best fit values of the model parameters i.e. $A, ~B,~ C, ~D$. The best fit values are $A = -0.96$, $B = 1.21$, $C = -0.62$ and $D=0.116$ respectively. We then differentiate Eq. \eqref{pfit} to get $P'^s_{fit}(z)$.

\begin{equation}
\label{derivativepfit}
\left(H_0\frac{d^2t}{dz^2}\right)^s=P'^s_{fit}(z)=B+2Cz+3Dz^2
\end{equation}
To calculate the  error associated with the model parameters, i.e $\sigma_A$, $\sigma_B$, $\sigma_C$ and $\sigma_D$,  we use the standard method as given in \cite{sivia}. We then use the  error propagation to calculate the error in the $P'^s_{fit}(z)$.

\begin{equation}
\label{psfiterror}
\sigma_{P'^s_{fit}}^2(z)=\sigma_B^2+4z^2\sigma_C^2+9z^4\sigma_D^2
\end{equation}
Figure \ref{d2tbydz2nps} shows the variation of reconstructed $H_0d^2t/dz^2$ with redshift and its $1\sigma$, $2\sigma$ \& $3\sigma$ error bands.

For dark energy models, we calculate $E'(z)$ from their corresponding $E(z)$. After substituting these expressions in Eq. \eqref{dtbydzderi}, we obtain $H_0d^2t/dz^2$ for the models.

\subsubsection{$\Lambda$CDM model}

We calculate the derivative of the dimensionless Hubble parameter $E(z)$ w.r.t. $z$ for the $\Lambda$CDM model,

\begin{equation}
\label{eq:hdslcdm}
E'_{\Lambda CDM}(z)=\frac{3 \Omega_{m0} (1+z)^2}{2E_{\small{\Lambda CDM}}(z)}
\end{equation}
Using values of $E'_{\Lambda CDM}(z)$ and $E_{\Lambda CDM}(z)$ in Eq. \eqref{dtbydzderi}, we find $H_0d^2t/dz^2$ i.e. $P'_{\Lambda CDM}(z)$.

\begin{equation}
\label{eq:pdslcdm}
P'_{\Lambda CDM}(z)=\frac{1}{(1+z)^2E_{\Lambda CDM}(z)}+\frac{E'_{\Lambda CDM}(z)}{(1+z)E^2_{\Lambda CDM}(z)}
\end{equation}

To calculate $\pm 1\sigma$ error in $H_0d^2t/dz^2$, we follow the same methodology as explained in Sec 3.1.1.

\begin{equation}
\label{sgpdslcdm}
\sigma_{P'+}=\sqrt{\sum_i \left(Max\left[\frac{dP'}{dx_{i}}\sigma_{x_{i+}}, \frac{-dP'}{dx_{i}}\sigma_{x_{i-}}\right]\right)^2 }
\end{equation}

\begin{equation}
\label{sgmdslcdm}
\sigma_{P'+}=\sqrt{\sum_i \left(Min\left[\frac{dP'}{dx_{i}}\sigma_{x_{i+}}, \frac{-dP'}{dx_{i}}\sigma_{x_{i-}}\right]\right)^2 }
\end{equation}

Similarly, to calculate the $P'$ for CPL, JBP \& FSLL, the expressions of $E'$ for each model are as follows:

\subsubsection{CPL Parametrization}

The derivative of $E(z)$ w.r.t. $z$ for the CPL parametrization is
\begin{multline*}
\label{hdscpl}
E'_{CPL}(z)=\frac{3}{2E_{CPL}(z)}\bigg[\Omega_{m0}(1+z)^2+(1-\Omega_{m0})(1+z)^{3(1+\omega_0+\omega_1)-1}\\
\exp\bigg(\frac{-3\omega_1z}{1+z} \bigg) \bigg((1+\omega_0+\omega_1)-\frac{\omega_1}{1+z} \bigg)\bigg]
\end{multline*}

\subsubsection{JBP Parametrization}

For this parametrization,
\begin{multline*}
E'_{JBP}(z)=\frac{3}{2E_{\small{JBP}}(z)}\bigg[\Omega_{m0}(1+z)^2+(1-\Omega_{m0})(1+z)^{3(1+\omega_2)-1}\\
\exp\bigg( \frac{3\omega_3z^2}{2(1+z)^2}\bigg)\bigg(\frac{\omega_3z}{(1+z)^2}+(1+\omega_2)\bigg)\bigg]
\end{multline*}

\subsubsection{FSLL Parametrization}

For the FSLL parametrization,
\begin{multline*}
E'_{FSLL}(z)=\frac{1}{2E_{FSLL}(z)}\bigg[3\Omega_{m0}(1+z)^2+(1-\Omega_{m0}){\textnormal{exp}}(1.5\omega_5{\textnormal{tan}^{-1}}(z))(1+z)^{\alpha}(1+z^2)^{0.75\omega_5}\\
\bigg(\frac{1.5\omega_5}{1+z^2}+\frac{\alpha}{(1+z)^{\alpha}}+\frac{1.5\omega_5z}{(1+z^2)} \bigg) \bigg]
\end{multline*}

where $\alpha=3(1+\omega_4-0.5\omega_5)$

\begin{figure}[ht]
\centering
\includegraphics[width=12.5cm, height=8.5cm]{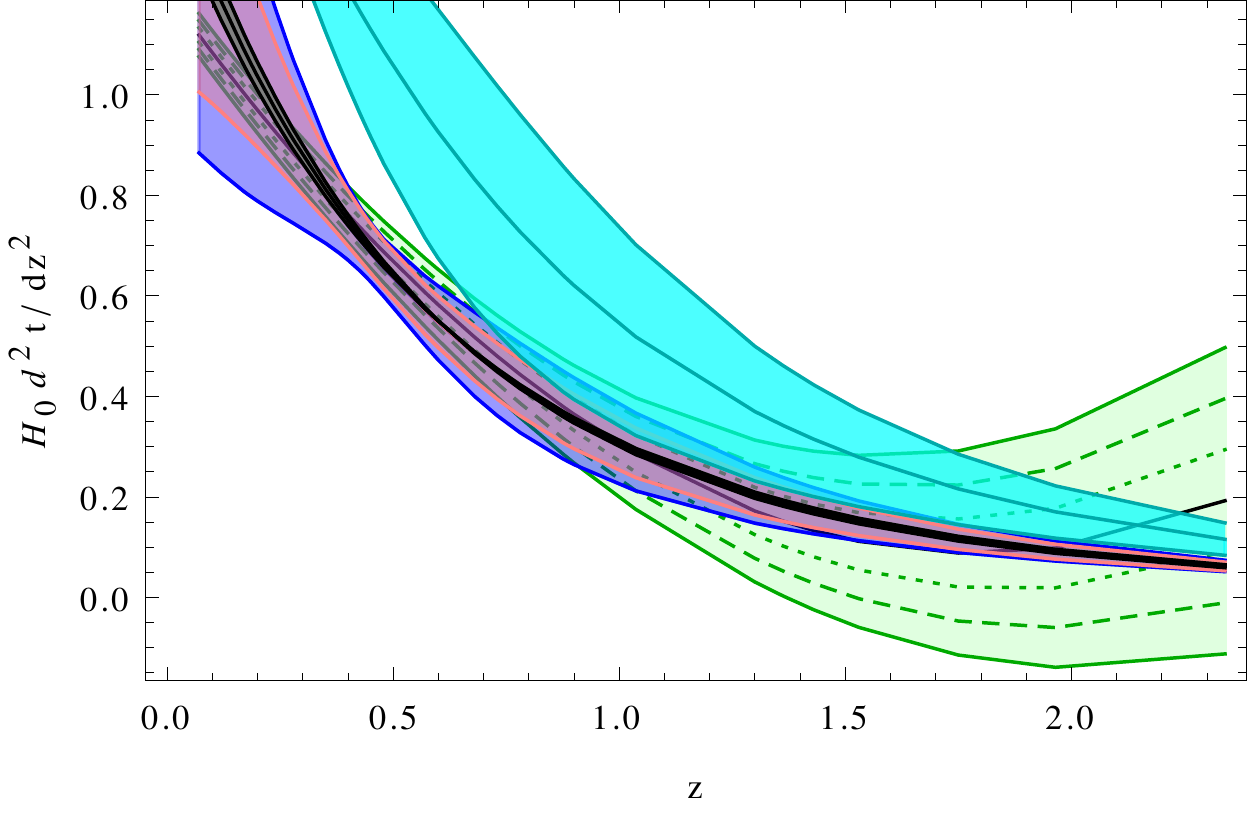}
\caption{\label{allmodeld2t}Variation of $H_0d^2t/dz^2$ with $z$ for all models and reconstructed (green) result. Black, pink, blue \& cyan represents $\Lambda$CDM, CPL, JBP \& FSLL respectively with their 3$\sigma$ error bands.}
\end{figure}

Figure \ref{allmodeld2t} shows $H_0d^2t/dz^2$ variation with $z$ for all models ($\Lambda$CDM-Black, CPL-Red, JBP-Blue, FSLL-Cyan) with their 3$\sigma$ error bands. The green shaded region shown by the dotted, dashed and continuous lines are the 1$\sigma$, 2$\sigma$ \& 3$\sigma$ confidence region of the reconstructed $H_0d^2t/dz^2$ respectively. The confidence regions of each model are obtained by using Eq. \eqref{sgpdslcdm} and \eqref{sgmdslcdm}.

\begin{figure}[ht]
\centering
\includegraphics[width=12.5cm, height=8.5cm]{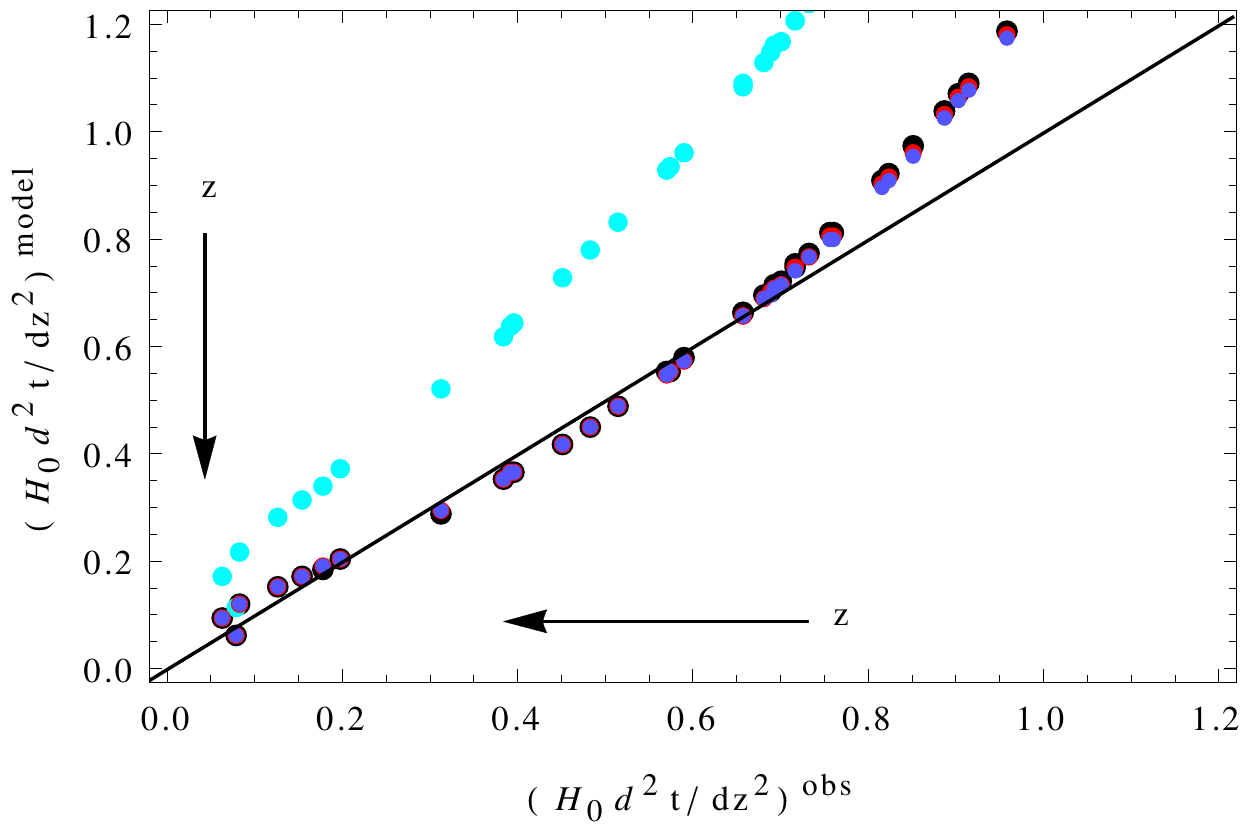}
\caption{\label{d2tobsmodel}Variation of $(H_0d^2t/dz^2)^{model}$ for $\Lambda$CDM, CPL, JBP \& FSLL model with the observed value of $(H_0d^2t/dz^2)^{obs}$ or $P'^s_{fit}$. Black, Red and blue colour represents $\Lambda$CDM, CPL and JBP model while cyan represents FSLL model. The arrow indicates the direction of increasing redshift.}
\end{figure}

Figure \ref{d2tobsmodel} shows that the FSLL model prediction deviates significantly from the smooth reconstructed graph, while the $\Lambda$CDM model deviates from the smooth results for z $\le0.3$.

\section{Discussion}

We compared the variation of $H_0dt/dz$ and its derivative for various dark energy models with the reconstructed results obtained by Non Parametric Smoothing. Our main conclusions from this study are as follows.
\begin{enumerate}

\item Models are compatible with NPS reconstruction whenever (3$\sigma$) uncertainty bands of NPS and the best fitted model itself overlap. The best fit line of CPL, JBP, FSLL parametrizations and the  $\Lambda$CDM model lie within the $3\sigma$ region of the reconstructed results obtained for the $H_0dt/dz$ formulation. Hence all four models seem to be in agreement with the smoothing results (See Figures \ref{dtbydzlcdmwidnps}, \ref{dtbydzcplwidnps}, \ref{dtbydzjbpwidnps}, \ref{dtbydzfsslwidnps}). But in the case of the phantom model, as shown in Figure \ref{dtbydzxcdmwidnps}, the best fit line deviates substantially from the reconstructed results at $z<1.5$. On the other hand, as expected, the EdS model shows clear disagreement with the smoothing result in the entire redshift range considered in this work (see Figure \ref{dtbydzEdSwidnps}).

\item We compared the model predictions with the observations for $H_0 dt/dz$ in a more compact form in Figure \ref{dtbydzobsnmodel}. The predictions of the Phantom and EdS model are significantly away from the diagonal line which acts as the benchmark for the model predictions to be consistent with the observations, while the other four model ($\Lambda$CDM, CPL, JBP and FSLL) predictions are scattered along the diagonal line.

\item As Jimenez and Loeb (2002) pointed out, the derivative of $H_0dt/dz$ tracks the equation of state directly. Following this idea, we further calculated $H_0d^2t/dz^2$ for the $\Lambda$CDM, CPL, JBP and FSLL models and compared with the corresponding reconstructed one, as shown in Figure \ref{allmodeld2t}. It is observed that above redshift ($z$) $0.3$, the best fit lines of the $\Lambda$CDM, CPL and JBP are consistent with the reconstructed one within 2$\sigma$. The predicted $3 \sigma$ confidence region of the $\Lambda$CDM model for $H_0d^2t/dz^2$ is narrow and consistent within the 3$\sigma$ bands of the reconstructed result in the redshift $ z > 0.3$. However, for CPL and JBP parametrizations, their 3$\sigma$ confidence region become wider compared to the reconstructed results at very low redshift. It looks that CPL parametrization is more compatible at $z>0.3$ than JBP parametrization. FSLL parametrization is not compatible with the reconstructed $H_0d^2t/dz^2$ within the 3$\sigma$ region.

\item Figure \ref{allmodeld2t} suggests that there is mismatch between the $\Lambda$CDM model predictions and the reconstructed $d^2t/dz^2$ using $H(z)$ observations at very low redshifts $z<0.3$. One possible reason could be the systematics associated with the data. Other effect could be the approximation scheme we adopted, i.e. fitting the $P(z)$ to a polynomial
and then differentiating it. If the polynomial was of a different order, the
result could be slightly different. However, the 3σ confidence region
of JBP and CPL parametrizations still accommodate the observations at
low redshifts. It might be an indication that evolving dark energy models explain the present acceleration of the universe better.

\item Figure \ref{d2tobsmodel} graphically demonstrates the strength of $H_0d^2t/dz^2$. $\Lambda$CDM, CPL and JBP models are consistent  with the reconstructed results  approximately at $z>0.3$. Below this  redshift ($z\leq 0.3$),  they not only deviate from the diagonal line but also start deviating among themselves. But the theoretical prediction of FSLL parametrization shows clear deviation from the black diagonal line. This shows that FSLL model of dark energy is not consistent with the reconstructed values of $H_0d^2t/dz^2$. So it can be seen that the degeneracy among the dark energy models is lifted to some extent with the help of $H_0d^2t/dz^2$. {\it{The point we would like to emphasize is that the derivative of $H_0dt/dz$ for various dark energy models shows different behaviour at low redshift along with their error bars. This feature is not visible in $H_0dt/dz$. Hence it seems possible to differentiate these dark energy models using the $H_0d^2t/dz^2$ method if we have enough data at low redshifts.}}
\end{enumerate}

It will be interesting to use differential age of galaxies as a tool on non-standard cosmology models like the Brane models or f(R) gravity in the future.

\acknowledgments
Authors are thankful to the anonymous referee for the suggestions. NR acknowledges financial support from the UGC Non-NET scheme (Govt. of India) and the facilities provided at IUCAA Resource Centre, Delhi University. AM thanks Research Council, University of Delhi, Delhi for providing support under R \& D scheme 2015-16. Authors are grateful to J. E. Gonzalez and Arman Shafieloo for useful discussion.


\begin{thebibliography}{99}

\bibitem{jimenez} Jimenez R. \& Loeb A.,  \emph{Constraining cosmological parameters based on relative galaxy ages}, \emph{ApJ} {\bf 573} (2002) 42 [arXiv:astro-ph/0106145]

\bibitem{riess} Riess A.G. et al., (Supernova Search Team Collaboration), \emph{Observational Evidence from Supernovae for an Accelerating Universe and a Cosmological Constant}, \emph{AJ}  {\bf 116} (1998) 1009 [arXiv:astro-ph/9805201].

\bibitem{perlmutter} Perlmutter S. et al., (Supernova Cosmology Project Collaboration), \emph{Measurements of Omega and Lambda from 42 High-Redshift Supernovae}, \emph{ApJ} {\bf 517} (1999) 565 [arXiv:astro-ph/9812133].

\bibitem{sean} Carroll S.M. \emph{The Cosmological Constant}, \emph{Living Reviews in Relativity} {\bf 04} (2001) 1 [arXiv:astro-ph/0004075].

\bibitem{frieman} Frieman J.A., Turner M.S. \& Huterer D., \emph{Dark Energy and the Accelerating Universe}, \emph{Annu. Rev. Astron. Astrophys.} {\bf 46} (2008) 385 [arXiv:0803.0982]

\bibitem{caldwell} Caldwell R. R. \& Kamionkowski M., \emph{The Physics of Cosmic Acceleration}, \emph{Annu. Rev. Nucl. Part. Sci.} {\bf 59} (2009) 397 [arXiv:0903.0866].

\bibitem{sne} Nesseris S. \& Perivolaropoulos  L., \emph{Tension and Systematics in the Gold06 SnIa Dataset}, \emph{JCAP} {\bf 02} (2007) 025  [arXiv:astro-ph/0612653]

\bibitem{sne1} Nesseris S. \& Perivolaropoulos L.,  \emph{Comparison of the Legacy and Gold SnIa Dataset Constraints on Dark Energy Models}, \emph{PRD} {\bf 72} (2005) 123519 [arXiv:astro-ph/0511040]

\bibitem{sne2} Lazkoz R., Nesseris S. \& Perivolaropoulos L., \emph{Exploring Cosmological Expansion Parametrizations with the Gold SnIa Dataset}, \emph{JCAP} {\bf 11} (2005) 010 [arXiv:astro-ph/0503230]

\bibitem{sne3} Nesseris S. \& Perivolaropoulos L., \emph{A comparison of cosmological models using recent supernova data}, \emph{PRD} {\bf 70} (2004) 043531 [arXiv:astro-ph/0401556]

\bibitem{dragan} Huterer D. \& Turner M.S., \emph{Constraining the properties of dark energy}, \emph{AIP Conf. Proc.} {\bf 586} (2001) 297 [arXiv:astro-ph/0103175]

\bibitem{linder} Linder E.V., \emph{Exploring the Expansion History of the Universe}, \emph{PRL}  {\bf 90} (2003) 091301 [arXiv:astro-ph/0208512]

\bibitem{george} Pantazis G., Nesseris S. \& Perivolaropoulos L., \emph{A Comparison of Thawing and Freezing Dark Energy Parametrizations}, \emph{PRD} {\bf 93} (2016) 103503 [arXiv:1603.02164]

\bibitem{jing} Qi J-Z, Zhang M-J \& Liu W-B, \emph{Testing dark energy models with $H(z)$ data} [arXiv:1606.00168]

\bibitem{celia} Escamilla-Rivera C., \emph{Status on Bidimensional Dark
Energy Parameterizations Using SNe Ia JLA and BAO Datasets}, \emph{Galaxies} {\bf 04} (2016) 08 [arXiv:1605.02702]

\bibitem{armangp} Shafieloo A., Alam U., Sahni V. \& Starobinsky A. A., \emph{Smoothing supernova data to reconstruct the expansion history
of the Universe and its age}, \emph{MNRAS}  {\bf 366} (2006) 1081     [arXiv:astro-ph/0505329]

\bibitem{arman} Shafieloo A., \emph{Model-independent reconstruction of the expansion history of the Universe and the properties of dark energy}, \emph{MNRAS}  {\bf 380} (2007) 1580 [arXiv:astro-ph/0703034]

\bibitem{puxun08} Wu P. and Yu H., \emph {Probing the cosmic acceleration history and the properties of dark energy from the ESSENCE supernova data with a model independent method}, \emph{JCAP} {\bf 02} (2008) 019 [arXiv:0802.2017]

\bibitem{chris2010} Shafieloo A. and Clarkson C., \emph{Model independent tests of the standard cosmological model},  \emph{PRD}  {\bf 81}  (2010) 083537 [arXiv:0911.4858]

\bibitem{arman2012} Shafieloo A., \emph{Crossing Statistic: Reconstructing the Expansion History of the Universe}, \emph{JCAP} {\bf 08} (2012) 002 [arXiv:1204.1109]

\bibitem{zhengli2015} Li Z.X. et al., \emph{Constructing a cosmological model-independent Hubble diagram of type Ia supernovae with cosmic chronometers}, \emph{PRD} {\bf 93} (2016) 043014 [arXiv:1504.03269]

\bibitem{data} Gonzalez J.E. et al., \emph{Non-parametric reconstruction of cosmological matter perturbations}, \emph{JCAP} {\bf 04} (2016) 016 [arXiv:1602.01015]

\bibitem{arman16} L'Huillier B. and Shafieloo A., \emph{Model-independent test of the FLRW metric, the flatness of the Universe, and non-local measurement of $H_0$}  [arXiv: 1606.06832]

\bibitem{jimenez2003} Jimenez R., \emph{The value of the equation of state of dark energy}, \emph{ApJ} {\bf 593} (2003) 622 [arXiv:astro-ph/0305368]

\bibitem{meliacc2015} Melia F. \& McClintock T.M., \emph{A Test of Cosmological Models using high-z Measurements of $H(z)$} [arXiv:1507.08279]

\bibitem{rafael2016} Rafael C.N. et al., \emph{New constraints on interacting dark energy from cosmic chronometers} [arXiv:1605.01712]

\bibitem{Ding2015} Ding X., Biesiada M., Cao S., Li Z.X. \& Zhu Z.-H., \emph{Is there evidence for dark energy evolution$?$
}, \emph{ApJL} {\bf 803} (2015) L22 [arXiv:1503.04923]

\bibitem[Zheng et al. (2016)]{Zheng2016}Li Z.X., Ding X., Biesiada M., Cao S. \& Zhu Z.-H., \emph{What are $Omh^2(z_1, z_2)$ and $Om(z_1, z_2)$ diagnostics telling us in light of $H(z)$ data?}, \emph{ApJ} {\bf 825}  (2016) 17 [arXiv:1604.07910]

\bibitem[Sahni, Shafieloo \& Starobinsky (2008)]{Sahni2008} Sahni V., Shafieloo A. \& Starobinsky A.A., \emph{Two new diagnostics of dark energy}, \emph{PRD} {\bf 78} (2008) 103502 [arXiv:0807.3548]

\bibitem{datameng} Meng et al., \emph{Utility of observational Hubble parameter data on dark energy evolution} [arXiv:1507.02517]

\bibitem{datamoresco} Moresco M. et al., \emph{A 6\% measurement of the Hubble parameter at $z\sim0.45:$ direct evidence of the epoch of cosmic re-acceleration}, \emph{JCAP} {\bf 05} (2016) 014 [arXiv:1601.01701].

\bibitem{baohz} Gaztanaga E., Cabre A. \& Hui L., \emph{Clustering of luminous red galaxies – IV. Baryon acoustic peak in the line-of-sight direction and a direct measurement of H(z)}, \emph{MNRAS} {\bf 399} (2009) 1663 [arXiv:0807.3551]

\bibitem{baohz1} Blake C. et al., \emph{The WiggleZ Dark Energy Survey: joint measurements of the expansion and growth history at $z < 1$}, \emph{MNRAS} {\bf 425} (2012) 405 [arXiv:1204.3674]

\bibitem{baohz2} Samushia L. et al., \emph{The clustering of galaxies in the SDSS-III DR9 Baryon Oscillation Spectroscopic Survey: testing deviations from $\Lambda$ and general relativity using anisotropic clustering of galaxies}, \emph{MNRAS} {\bf 429} (2013) 1514 [arXiv:1206.5309]

\bibitem{baohz3} Xu X. et al., \emph{Measuring $D_A$ and H at $z=0.35$ from the SDSS DR7 LRGs using baryon acoustic oscillations }, \emph{MNRAS} {\bf 431} (2013) 2834 [arXiv:1206.6732]

\bibitem{baohz4} Delubac T. et al., \emph{Baryon acoustic oscillations in the Ly$\alpha$ forest of BOSS DR11 quasars}, \emph{A} \& \emph{A} {\bf 574} (2015) A59 [arXiv:1404.1801]

\bibitem{planck} Planck Collaboration: Ade P.A.R. et al., \emph{Planck 2015 results. XIII Cosmological Parameters}, \emph{A} \& \emph{A} {\bf 594} (2016) A13 [arXiv:1502.01589]

\bibitem{demodels} Copeland E.J., Sami M. \& Tsujikawa S.,  \emph{Dynamics of dark energy}, \emph{IJMPD} {\bf 15} (2006) 1753 [arXiv:hep-th/0603057]

\bibitem{demodels1} Yoo J. \& Watanabe Y., \emph{Theoretical models of dark energy}, \emph{IJMPD} {\bf 21} (2012) 1230002 [arXiv:1212.4726]

\bibitem{error1} Lazkoz R., Montiel A. \& Salzano V., \emph{First cosmological constraints on the superfluid Chaplygin gas model}, \emph{PRD} {\bf 86} (2012) 103535 [arXiv:1211.3681]

\bibitem{error2} Qi J-Z et al., \emph{Transient acceleration in f(T) gravity}, \emph{RAA} {\bf 16} (2016) 002 [arXiv:1403.7287]

\bibitem{cpl1} Chevallier M. and Polarski D., \emph{Accelerating universes with scaling dark matter}, \emph{IJMPD} {\bf 10} (2001) 213 [arXiv:gr-qc/0009008]

\bibitem{cpl2} Linder E. V., \emph{Exploring the expansion history of the Universe}, \emph{PRL} {\bf 90} (2003) 091301 [arXiv:astro-ph/0208512]

\bibitem{wang} Wang S. et al., \emph{A Comprehensive investigation on the slowing down of cosmic acceleration}, \emph{ApJ} {\bf 821} (2016) 60
[arXiv:1509.03461]

\bibitem{jbp} Jassal H. K., Bagla,J. S. and Padmanabhan T., \emph{WMAP constraints on low redshift evolution of dark energy}, \emph{MNRAS} {\bf 356} (2005) L11 [arXiv:astro-ph/0404378]

\bibitem{fssl} Feng C. J., Shen X. Y., Li P. \& Li Y. Z., \emph{A new class of parametrization for dark energy without divergence}, \emph{JCAP} {\bf 09} (2012) 023 [arXiv:1206.0063]

\bibitem{sivia} Sivia D.S.  \& Skilling J. \emph{Data Analysis: A Bayesian Tutorial-} II Edition, Oxford University Press, New York (2006)

\begin{table}[H]
\label{Observed data table}
\caption{Hubble data}
\centering
\begin{tabular}{|c|c|c|c|}
\hline\hline
$z_i$ 			& $H^{ob}(z_i)$ & $\sigma_H^{ob}(z_i)$ & Technique \\
\hline
0.0708	&	69		& 	19.68	 & 	DA\\
0.09	&	69		& 	12		 & 	DA\\
0.12	&	68.6 	&	26.2	 &	DA\\
0.17	&	83		&	8		 &	DA\\
0.179	&	75		&	4		 &	DA\\
0.199	&	75		&	5   	& DA\\
0.20	&	72.9	&	29.6	& DA\\
0.27	&	77		&	14	&	DA\\
0.28	&	88.8	&	36.6	&	DA\\
0.352	&	83		&	14	&	DA\\
0.3802	&	83		&	13.5	&	DA\\
0.4		&	95		&	17	&	DA\\
0.4004	&	77		&	10.2	&	DA\\
0.4247	&	87.1	&	11.2	&	DA\\
0.4497	&	92.8	&	12.9	&	DA\\
0.4783	&	80.9	&	9	&	DA\\
0.48	&	97		&	62	&	DA\\
0.593	&	104		&		13	&	DA\\
0.68	&	92		&	8	&	DA\\
0.781	&	105	&		12	&	DA\\
0.875	&	125	&	17&DA\\
0.88	&	90		&	40&DA\\
0.9&		117	&		23&DA\\
1.037&	154		&	20&DA\\
1.3		&168	&		17&DA\\
1.363&	160		&	33.6&DA\\
1.43	&177	&		18&DA\\
1.53	&140	&		14&DA\\
1.75	&202	&		40&DA\\
1.965&	186.5	&	50.4	&DA\\
0.240&	79.69	&	2.65&BAO\\
0.35	&84.4	&	7&BAO\\
0.43	&86.45	&	3.68&BAO\\
0.44	&82.6	&	7.8&BAO\\
0.57	&92.4	&	4.5&BAO\\
0.6		&87.9	&	6.1&BAO\\
0.73	&97.3	&	7&BAO\\
2.34	&222	& 7&BAO\\
\hline
\end{tabular}
\end{table}

Here DA represents Differential Age approach.


\end{thebibliography}
\end{document}